\documentclass[sigconf]{acmart}
\usepackage{balance}
\usepackage{booktabs}
\usepackage{enumitem}
\usepackage{xcolor}
\definecolor{codegreen}{rgb}{0,0.6,0}
\definecolor{codegray}{rgb}{0.5,0.5,0.5}
\definecolor{codepurple}{rgb}{0.58,0,0.82}
\definecolor{backcolour}{rgb}{0.95,0.95,0.92}
\definecolor{dkgreen}{rgb}{0,0.6,0}    
\usepackage{algorithmic}
\usepackage[linesnumbered,ruled,vlined]{algorithm2e}

\SetCommentSty{mycommfont}

\usepackage{xspace}
\usepackage{multirow}
\usepackage{pifont}
\usepackage{subfigure}

\newcommand{\tool}{\emph{TransRepair}\xspace}

\newcommand{\revise}[1][\textcolor{black}]{#1}

\newcommand{\ie}{\textit{i.e.}\xspace}
\newcommand{\eg}{\textit{e.g.}\xspace}
\newcommand{\etal}{\textit{et al.}\xspace}

\AtBeginDocument{%
  \providecommand\BibTeX{{%
    \normalfont B\kern-0.5em{\scshape i\kern-0.25em b}\kern-0.8em\TeX}}}

\copyrightyear{2022}
\acmYear{2022}
\setcopyright{rightsretained}
\acmConference[ASE '22]{37th IEEE/ACM International Conference on Automated Software Engineering}{October 10--14, 2022}{Rochester, MI, USA}
\acmBooktitle{37th IEEE/ACM International Conference on Automated Software Engineering (ASE '22), October 10--14, 2022, Rochester, MI, USA}
\acmDOI{10.1145/3551349.3560422}
\acmISBN{978-1-4503-9475-8/22/10}
\begin{document}

\title{\tool: Context-aware Program Repair \\ for Compilation Errors}
\renewcommand{\shorttitle}{\tool: Context-aware Program Repair for Compilation Errors}

\author{Xueyang Li}
\authornotemark[1]
\affiliation{%
  \institution{SKLOIS, IIE, CAS}
  \institution{School of Cybersecurity, UCAS}
  \country{China}
}
\author{Shangqing Liu}
\authornote{Both authors contributed equally to this research.}
\affiliation{%
  \institution{Nanyang Technological University}
  \country{Singapore}
}
\author{Ruitao Feng}
\affiliation{%
  \institution{University of New South Wales}
  \country{Australia}
}
\author{Guozhu Meng}
\authornote{Corresponding author.}
\affiliation{%
  \institution{SKLOIS, IIE, CAS}
  \institution{School of Cybersecurity, UCAS}
  \country{China}
}
\author{Xiaofei Xie}
\affiliation{%
  \institution{Singapore Management University}
  \country{Singapore}
}
\author{Kai Chen}
\affiliation{%
  \institution{SKLOIS, IIE, CAS}
  \institution{School of Cybersecurity, UCAS}
  \institution{BAAI}
  \country{China}
}
\author{Yang Liu}
\affiliation{%
  \institution{Nanyang Technological University}
  \country{Singapore}
}

\settopmatter{printacmref=True}



\begin{abstract}
Automatically fixing compilation errors can greatly raise the productivity of software development, by guiding the novice or AI programmers to write and debug code. 
Recently, learning-based program repair has gained extensive attention and became the state-of-the-art in practice.
But it still leaves plenty of space for improvement. 
In this paper, we propose an end-to-end solution~\tool to locate the error lines and create the correct substitute for a C program simultaneously. 
Superior to the counterpart, our approach takes into account the context of erroneous code and diagnostic compilation feedback. Then we devise a Transformer-based neural network to learn the ways of repair from the erroneous code as well as its context and the diagnostic feedback.
To increase the effectiveness of \tool, we summarize 5 types and 74 fine-grained sub-types of compilations errors from two real-world program datasets and the Internet.
Then a program corruption technique is developed to synthesize a large dataset with 1,821,275 erroneous C programs. 
Through the extensive experiments, we demonstrate that \tool outperforms the state-of-the-art in both single repair accuracy and full repair accuracy. Further analysis sheds light on the strengths and weaknesses in the contemporary solutions for future improvement.
\end{abstract}




\maketitle

\section{Introduction}\label{sec:intro}
Automated program repair, which aims at fixing the underlying errors in a program, plays a critical role in the software development cycle. 
Generally, it can be roughly categorized into program logical error fixing and compilation error fixing. Compared with the widespread attention on repairing program logical errors~\cite{li_dlfix_2020, seqr, lutellier_coconut_2020, cure}, the compilation error fixing has just gotten into the horizon of researchers in the past few years~\cite{gupta2017deepfix, yasunaga2020graph, ahmed2018compilation}. 
Besides raising the productivity of software development, it can also facilitate the AI programming, such as code generation~\cite{copilot,plotcoder2021acl} and binary decompilation~\cite{coda2019nips,dec2019arxiv}. 
Recent research shows that AI programmers may produce lots of erroneous code (including compilation errors) as human novice programmers did~\cite{code2022chi}. However, it is non-trivial yet to automatically fix compilation errors in an undocumented program~\cite{DBLP:conf/iticse/DennyLT12}. 
Moreover, the error messages returned by a compiler may be obscure and cryptic considering the compiler is evolving with new features and optimization techniques~\cite{DBLP:journals/ahci/Traver10}. 
\revise{As a consequence, it is desired and beneficial that the program with compilation errors can be automatically repaired to raise programming productivity and prompt AI programming.}

\begin{figure}[t]
\centering
\centering
     \includegraphics[width=1\linewidth]{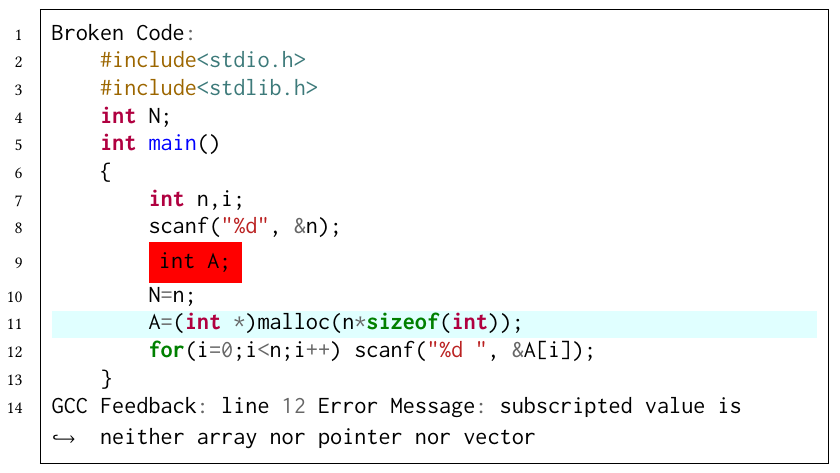}
\caption{The broken code with its compiler message.}
\label{fig:intro_example}
\vspace{-5mm}
\end{figure}

Automated program repair for compilation errors is a far-from-settled problem. 
Prior studies~\cite{gupta2017deepfix, bhatia2016automated, santos2018syntax, ahmed2018compilation} directly utilized RNN-based encoder-decoder framework to take as input the broken program to generate the exact fix. However, the selected model architecture has the limited learning capacity and drawbacks such as RNNs struggle with long-range dependencies in a sequence. Furthermore, other studies~\cite{ahmed2021synfix, yasunaga2020graph, mesbah2019deepdelta} have demonstrated that the compiler diagnostic feedback is valuable to improve the accuracy. For example, DrRepair~\cite{yasunaga2020graph} proposed to construct the program-feedback graph by connecting same identifiers in source code and symbols (\eg, identifiers, types, operators) in the compiler feedback to encode the semantic correspondence and further utilized graph attention network to capture relations between program and message to fix the broken program. DrRepair has achieved the state-of-the-art performance and outperforms previous approaches that ignore the compiler feedback significantly. However, through our in-depth analysis of the feedback produced by the compiler, we find that the correspondence between the location of the broken code and the error message is not completely accurate. A simple example is illustrated in Figure~\ref{fig:intro_example}.
It shows that the feedback produced by GCC compiler consists of the reported line number (\ie, line 12 in Figure~\ref{fig:intro_example}) and the error messages. 
The root cause is at line 9 and the identifier $A$ should be declared as a pointer type (\ie, ``int A'' $\rightarrow$ ``int $*$A''). However, the feedback produced by GCC depicts that there is an error at line 12. The location of the root cause in the broken program and the line number produced in the feedback are mismatched, which demonstrates that the error message fails to reveal the reason of this error. Hence, the graph constructed 
based on the feedback may not capture the essence of errors. Furthermore, in Figure~\ref{fig:intro_example}, we also find that there is no symbol existing in the feedback and the program-feedback graph cannot be constructed. Finally, the context (highlighted in blue of Figure~\ref{fig:intro_example}) can infer that the identifier A is a pointer rather than an integer, but this part of context information is ignored in current works.

On the other hand, high quality training data is demanding for learning-based program repair~\cite{icse2022code}. There are two open-source datasets with compilation errors of C programming language (\ie, DeepFix~\cite{gupta2017deepfix} and TRACER~\cite{ahmed2018compilation}). The DeepFix dataset contains 37,415 correct programs and 6,971 broken programs, which fail to pass the compilation and TRACER contains 21,994 single-line error programs\footnote{The exact number is mismatched with the reported number in the original paper~\cite{ahmed2018compilation}, since we filter out some obvious error samples.}.
Although the dataset is further augmented~\cite{yasunaga2020graph} by a program corruption approach, the synthesized code is limited in error types so that the repair performance will be greatly degraded in front of arbitrary errors in reality.
Additionally, the data for training a repair model is not yet extensively evaluated, so it is unclear what types of errors cannot be well learned and the underlying cause.

To address the aforementioned challenges, in this study, we propose a context-aware program repair technique to fix compilation errors. 
To enrich the diversity of the broken programs, we conduct a comprehensive analysis on compilation errors from two real-world programs (\ie, DeepFix and TRACER) and relevant questions in StackOverflow. 
We summarize these common compilation errors and obtain \textbf{74} compilation errors in terms of syntax and semantics. We further classify these errors in \textbf{5} different groups. 
We propose fine-grained perturbation strategies for each type of tokens in a program, and develop an automated approach to break programs with specific errors. 
In such a manner, we synthesize a dataset with \textbf{1,821,275} broken programs in line with the real error scenario.
We further devise a Transformer-based program repair model (\ie, \tool) that takes as input each line of a broken program, the context for each line of statements and the error message to locate the errors and then fix them. A pointer mechanism is incorporated into the model that proves to be effective in solving errors involved with \emph{out-of-vocabulary} code tokens. 
The extensive experiments on two open-source dataset DeepFix and TRACER have demonstrated that \tool outperforms current state-of-the-art DrRepair in repair accuracy by 4.66\% and 5.7\% on DeepFix and TRACER, respectively. The ablation studies for both model components and training data reveal the importance in lifting the repair efficacy.
The result analysis concludes that our approach performs the best in fixing ``statement'' errors and gains more advantages for ``type mismatch'' and ``variable declaration'' errors compared to DrRepair.

\noindent\textbf{Contributions.} We summarize the main contributions as follows:
\begin{itemize}[leftmargin=*]
    \item We empirically analyze the common compilation errors from two public datasets and StackOverflow, concluding 74 concrete patterns of compilation errors and 5 categories. Based on that, we further design a number of fine-grained perturbation strategies to create a dataset of diverse broken problems. 
    \item We propose a Transformer-based repair model, which takes each line of a broken program, its context and error messages as input to locate and repair the erroneous code. According to the best of our knowledge, we are the first to consider the context information for repairing the compilation errors.
    \item The extensive experiments on two open-source datasets demonstrate that \tool outperforms the state-of-the-art in both single repair and full repair. Moreover, the ablation and failure case studies identify the inherent advantages and limits in light of different types of errors.  
\end{itemize}

More details about code, model and experimental results can be accessed from \cite{website} to benefit the academia and industry. The rest of this paper is organized as follows. Section~\ref{sec:overview} presents an overview of our approach. Section~\ref{sec:synthesis} introduces the data synthesis to construct a corrupted dataset. Section~\ref{sec:parsing} and Section~\ref{sec:repair} are the detailed presentation of data parsing and model design. We introduce the experimental setup and analyze experimental results in Section~\ref{sec:eval} and Section~\ref{sec:results} respectively. Section~\ref{sec:discuz} details the threats to validity of our work, followed by the related work in Section~\ref{sec:related}. We conclude our paper in Section~\ref{sec:concl}.

\begin{figure*}[t]
     \centering
     \includegraphics[width=1\linewidth]{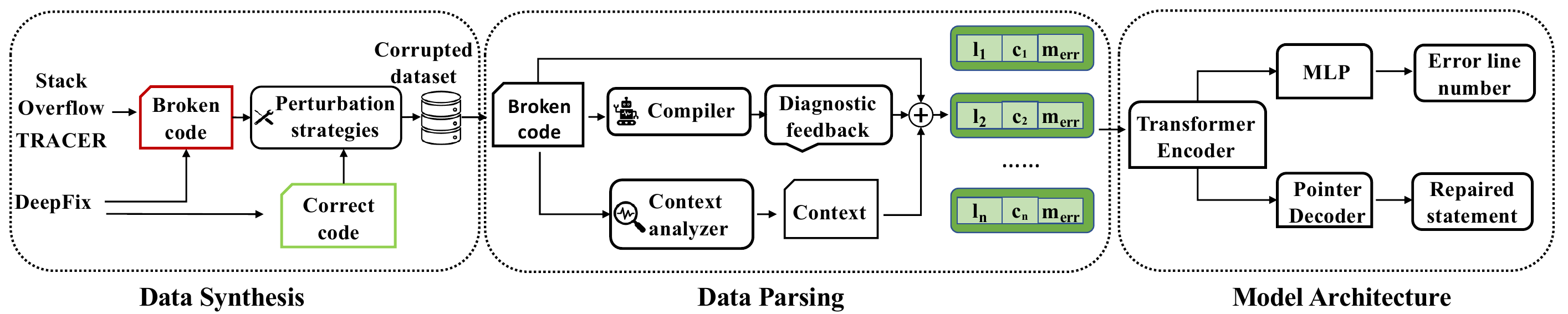}
     \caption{The overview of \tool}
     \label{fig:overview}
     \vspace{-6mm}
\end{figure*}

\section{System Overview}\label{sec:overview}
In this section, we first formulate the research problem, then provide an overview of our approach. 
\subsection{Problem Formulation}
Following the existing works~\cite{ahmed2021synfix, yasunaga2020graph, mesbah2019deepdelta}, \tool aims at repairing the program compilation errors by learning the program semantics through deep learning techniques. Formally, given a broken program $p$ from a dataset $D$ (\ie, $p \in D$), where $p = (l_1,l_2,...,l_n)$, $n$ is the total number of lines in $p$. Its diagnostic feedback provided by a compiler is defined as a list of $(i_{\mathrm{err}}, m_{\mathrm{err}}), $
where $i_{\mathrm{err}}$ is the reported line number, and $m_{\mathrm{err}}$ is the error message. Since the line number in the diagnostic feedback may not match the line of the root cause in a broken program (shown in Figure~\ref{fig:intro_example}), the goal of \tool is to learn a function $f$ from the dataset $D$ that takes $(p, i_{\mathrm{err}}, m_{\mathrm{err}})$ as input and identifies the location $k$ of the erroneous code $l_k$ where $k \in \{1,...,n\}$, and a repaired version of this statement (\ie, $l'_k$). The formulation can be expressed as $l'_k=f(p, i_{\mathrm{err}}, m_{\mathrm{err}})$. 

\subsection{Approach Overview}
Figure~\ref{fig:overview} presents the overview of our approach and it consists of three sequential modules--\emph{data synthesis}, \emph{data parsing} and \emph{model architecture}. In the data synthesis, we first empirically summarize the common compilation errors from multiple error sources including DeepFix, TRACER and a self-curated dataset from StackOverflow. We further design a set of perturbation strategies based on the summarized compilation errors to corrupt the correct programs from DeepFix and construct a new high-quality dataset $D$ that is in line with the real scenario. For each broken program $p$ in the constructed dataset, we compile it to obtain the diagnostic feedback (\ie, $(i_{\mathrm{err}}, m_{\mathrm{err}})$) provided by the compiler. Furthermore, we design a context analyzer to extract the context of each line of code to facilitate learning the context by the model. We take each line $l_i$, its context $c_i$ as well as the diagnostic feedback $(i_{\mathrm{err}}, m_{\mathrm{err}})$ as the input of the Transformer encoder to learn vector representations. We further apply a fully-connected feedforward network (MLP) to locate the line with error, and a pointer-based Transformer decoder to generate a repair for the error code. 

\section{Data Synthesis}\label{sec:synthesis}
In this section, we introduce our data synthesis module that aims at corrupting the correct program by the summarized perturbation strategies to construct a high-quality corrupted dataset in line with the real scenario. 

\begin{table*}[]
\small
\caption{The analysis of common compiler errors from DeepFix, TRACER and StackOverflow as well as the correspond program perturbation operation, which consists of the operand to change and operations.}
\label{tbl-statistics}
\begin{tabular}{cccccccccc}
\toprule
\multirow{2}{*}{\textbf{Error}}  & \multirow{2}{*}{\textbf{Type}}        & \multicolumn{4}{c}{\textbf{Statistics}}    & \multirow{2}{*}{\textbf{Operand}}  & \multicolumn{3}{c}{\textbf{Operation}}  \\ \cline{3-6} \cline{8-10}

&   & \multicolumn{1}{c}{\textbf{DeepFix}} & \multicolumn{1}{c}{\textbf{StackO.}} & \multicolumn{1}{c}{\textbf{Trace}} & \textbf{Avg.} &          &              \multicolumn{1}{c}{\textbf{ADD}} & \multicolumn{1}{c}{\textbf{DEL}} & \textbf{REP}         \\ \midrule

\multirow{2}{*}{Syntax} & structure (struct)                    & \multicolumn{1}{c}{56.51\%}        & \multicolumn{1}{c}{14.00\%}              & \multicolumn{1}{c}{19.68\%}       &     21.28\%&  punctuator &    \multicolumn{1}{c}{\checkmark}    & \multicolumn{1}{c}{\checkmark}         &   \checkmark             \\

     & statement (stmt)                    & \multicolumn{1}{c}{69.40\%}        & \multicolumn{1}{c}{34.00\%}              & \multicolumn{1}{c}{69.04\%}  &   51.52\%   &   keywords/operator/variable type/name  &   \multicolumn{1}{c}{\checkmark}    & \multicolumn{1}{c}{\checkmark}  &   \checkmark   \\ \midrule
\multirow{3}{*}{Semantic}   & variable declaration (decl) & \multicolumn{1}{c}{52.85\%}        & \multicolumn{1}{c}{39.50\%}              & \multicolumn{1}{c}{20.88\%}       &  21.43\%  &    variable type/name     &       \multicolumn{1}{c}{\checkmark}    & \multicolumn{1}{c}{}         &       \\
                        & type mismatch (tm)                & \multicolumn{1}{c}{2.95\%}        & \multicolumn{1}{c}{4.00\%}              & \multicolumn{1}{c}{2.87\%}       &    2.17\%  &   variable type/name    &      \multicolumn{1}{c}{\checkmark}    & \multicolumn{1}{c}{\checkmark}         &              \\
                        & identifier misuse (im)            & \multicolumn{1}{c}{2.61\%}        & \multicolumn{1}{c}{10.50\%}              &  \multicolumn{1}{c}{5.47\%}       &   3.60\%   &    operator/variable name   & \multicolumn{1}{c}{\checkmark}    & \multicolumn{1}{c}{}         &    \checkmark        \\ \bottomrule
\end{tabular}
\vspace{-3mm}
\end{table*}

\subsection{Taxonomy of Compilation Errors}\label{sec:taxonomy}
High quality data (\eg, large number, good diversity and accurate error triage) makes a model better learn the repair rules. The study~\cite{yasunaga2020graph} summarizes common compilation errors for Java, C and C++ programming languages from DeepDelta~\cite{mesbah2019deepdelta}, DeepFix~\cite{gupta2017deepfix} and SPoC~\cite{kulal2019spoc} respectively. 
Then five types of errors are specified as well as the corresponding corruption rules for broken code synthesis.
However, as we observe, there are more types of compilation errors that appear in reality but not in the their datasets.


In this study, we construct our own dataset by manually analyzing 6,971 erroneous programs in DeepFix and 21,994 programs in TRACER. Furthermore, we conduct an intensive search in StackOverflow to include more diverse errors. Specifically, to obtain a collection of compilation errors, we retrieve the data on StackOverflow with the keywords ``[syntax-error] [c]'' or ``[compile-error] [c]'' and get 200 questions 
\revise{ranked by ``Highest score''} ~\footnote{The queried results are as of April, 2022.}.  
\revise{All the programs as well as their error messages in StackOverflow are enclosed into our dataset.}

\revise{\textbf{Manual analysis.} We recruited four experts, all of whom have more than five years of programming experience, to analyze the collected program errors from DeepFix, TRACER and StackOverflow. 
First, we normalize the error messages by removing the specific information such as identifier name and line number, and group them with the same normalized messages into distinct clusters. Then, we spend about six man months to identify the type of errors, and whether an error message is accurate, for example, in revealing the causes of code errors. Specifically, we divide these clusters into four analysis tasks and assign one expert with two of them. Every error message is analyzed by two experts for cross validation. If a disagreement occurs, a third expert will be involved to make the final decision.}

\revise{
The compiler usually conducts the syntax analysis and semantic analysis to ensure the correction of a program. 
For example, the mistakenly spell of reserved words can incur a syntax error and using a variable without declaration produces a semantic error.
As aforementioned, we manually analyze the collected erroneous programs and distill a list of 74 error patterns in total. 
As shown in Table~\ref{tbl-statistics}, we further cluster these patterns into five categories within the syntax and semantic analysis phases. 
This taxonomy is built mainly based on the principles of compiler~\cite{compiler} and the analysis objects in each phase.
In particular, a compiler will check whether the program complies with the context-free grammar of C in syntax analysis and produce syntax errors if failed.
As observed in the dataset, there are two types of errors-structure error and statement error, significantly varying in influence scope and repair strategies. 
Structure error defines the misuse or absence of delimiter(s) (\eg, ``\{'', ``\}'', ``;'') in a statement or a block. It may propagate the influence to the entire program when a brace, for example, is missing. 
On the contrary, statement errors are caused due to the mistaken tokens in labeled statement, expression statement, selection statement or iteration statement, and the error influence is often confined in a single line. For example, for a correct expression statement ``a = a + 1'', if ``1'' is missing, the expression becomes ``a = a +'', which can definitely cause an error with single-line influence. 
In semantic analysis, the compiler will build the semantics for the constructs of code as well as their relations in between.
Therefore, errors are identified specific to the concrete semantic analysis tasks, such as \emph{scope resolution} and \emph{type checking}.
Here we refine semantic errors into three classes, namely ``variable declaration'', ``type mismatch'' and ``identifier misuse''. The ``variable declaration'' represents the use before the variable is declared. The error of ``type mismatch'' defines the mismatch of the type or the number of formal parameters of a function. For example, given a function ``$f(a, b)$'' that allows the invocation with two arguments, however, it is fed with three arguments, \eg, ``$f(a, b, c)$'', inducing such errors. 
As for ``identifier misuse'', for example, a variable is declared as an Integer, so that it cannot be used as a pointer like ``int a; a->t=0;''.} 

The statistics of these types of errors in the datasets of DeepFix, StackOverflow and TRACER is also presented in Table~\ref{tbl-statistics}. As a program may have multiple types of compilation errors, the total ratio of each dataset may exceed 100\%. We observe that the distribution of compilation errors are very different across the datasets. 
Generally, the structure, statement and variable declaration account for the vast majority in the datasets.

\vspace{-3mm}
\subsection{Broken Code Synthesis}
To prepare the broken programs with the aforementioned errors, we devise a specific perturbation method to corrupt the correct programs from DeepFix. 
The code corruption is conducted token-wised, that is, we make changes to a certain code token to produce an error. There are basically three operations in the course of perturbation--\emph{ADD} is to add one token; \emph{DEL} means to remove one token, and; \emph{REP} works as replacing a token with another one. As such, the synthesis of broken code proceeds in the following steps. 

\noindent\textbf{Step 1.} Given a program, we construct its abstract syntax tree (AST) and identify all the tokens in code, as well as the type of tokens. 

\noindent\textbf{Step 2.} Configure the corruption procedure by specifying the number of errors made to the code, and the type of errors. Here we create at most five errors for each program, in order to  enable the repairer to be able to fix the code with multiple errors. 

\noindent\textbf{Step 3.} Make the errors specified in the previous step. For each error, 
\revise{ we first conduct a global analysis of the target code, select the candidate variable names or symbols for replacement according to the corruption rules, and finally select one of them as the operand.} For example, to generate a ``statement'' error, we can take the keyword, operator, variable type or name in AST as the operand, and perform one of three operations (\ie, add, delete and replace). Table~\ref{tbl-statistics} shows the details for perturbation strategies. Noted that, 
\revise{when the operation type is REP, we will first find the tokens in the context based on the specific error type, and then randomly select one from them.}

\begin{figure}
	\centering
	\includegraphics[width=0.45\textwidth]{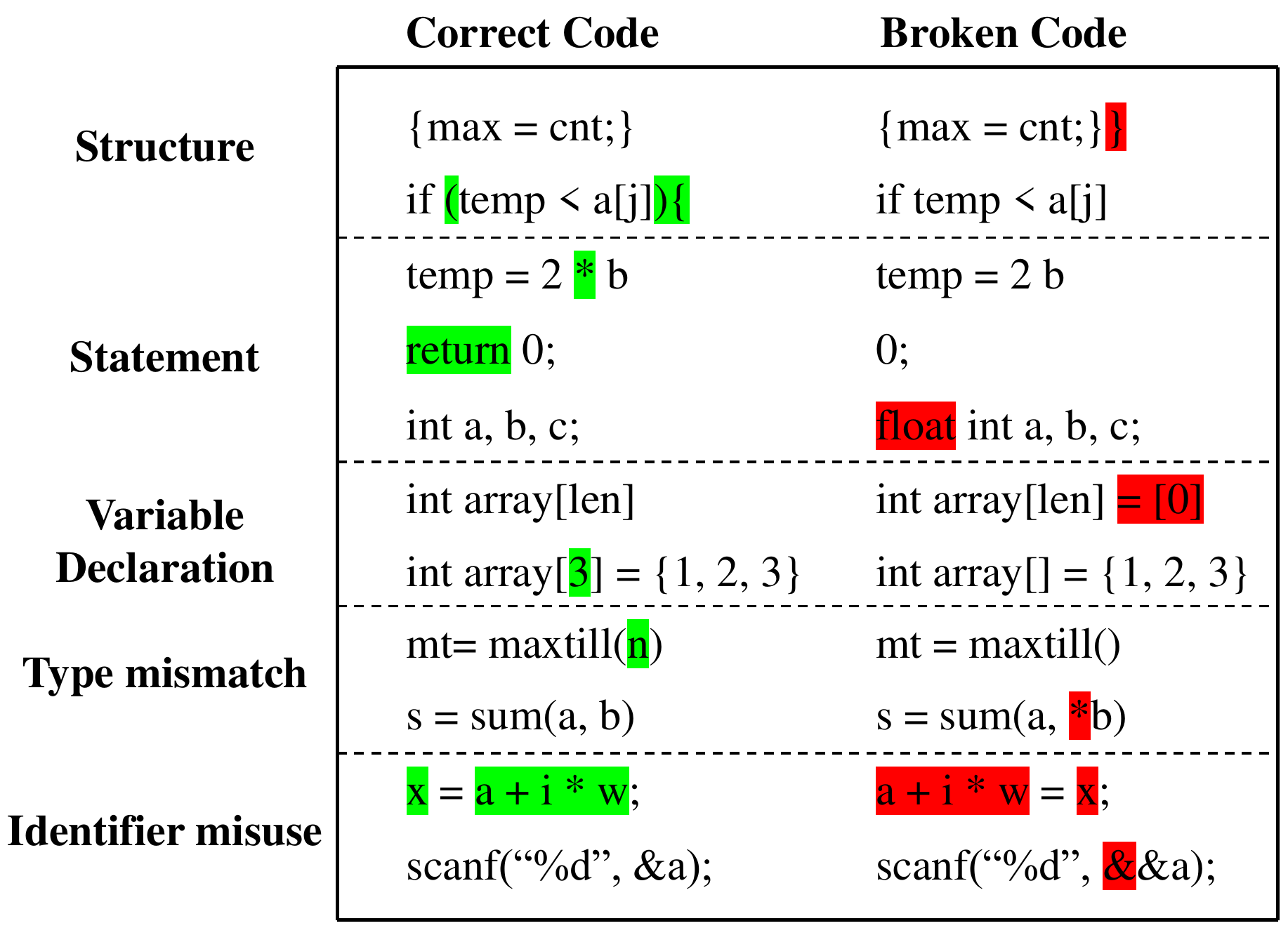}
	\caption{Examples of synthesized broken code}\label{fig:example}
	\vspace{-6mm}
\end{figure}

The following part presents how to corrupt programs to generate specific errors.
\begin{itemize}[leftmargin=*]
\item {\bf Structure,} which randomly adds, deletes or replaces an punctuator such as ``,.;(){}[]'' at the position of punctuator. 
\item {\bf Statement,} which randomly adds, deletes or replaces a keyword/operator/variable type/variable name at any statement if it has such features.
\item {\bf Variable declaration,} which adds a variable type or variable name at the variable declaration/usage statement to corrupt a program.
\item {\bf Type mismatch,} which randomly adds or deletes a variable type or variable name in the argument list of the function invocation. 
\item {\bf Identifier misuse,} which randomly adds or deletes an operator or variable name at the declaration statement. 
\end{itemize}
We present the perturbation strategies with some examples in Figure~\ref{fig:example} for better illustration. 
For each correct program, we repeatedly conduct the code synthesis procedure for 50 times to generate different broken programs and construct a new dataset $D$.
\revise{Additionally, compared with \cite{yasunaga2020graph}, our rules for perturbation are summarized from multiple program sources, which are in line with the real-world programming errors. All the above enables us to prepare a better training set for program repair learning. As a consequence, it makes the model to learn more diverse and comprehensive compilation errors, and achieve better repair efficiency as shown in Section~\ref{sec:eval:comparison}. }

\section{Data Parsing}\label{sec:parsing}
Through Section~\ref{sec:synthesis}, we can construct a new dataset $D$, where the program $p$ in this dataset (\ie, $p \in D$) has some compilation errors. In this section, we introduce the module of diagnostic feedback extraction and the context extraction. 

\vspace{-2mm}
\subsection{Extraction of Diagnostic Feedback}
The previous works~\cite{ahmed2021synfix, yasunaga2020graph, mesbah2019deepdelta} have confirmed that the diagnostic feedback could improve the localization and repair accuracy greatly. Hence, we also incorporate it in \tool. 
Specifically, since a broken program may consist of multiple errors, making it possible for the compiler to return multiple error messages,
\revise{we take into account all these errors in the training phase. But in the course of the validation phase, we perform an iterative process to repair each error successively. In each iteration, we use the first error which consists of the reported line number $i_{\mathrm{err}}$ and the error message $m_{\mathrm{err}}$ as \cite{yasunaga2020graph}.}
Furthermore, we replace the function name, variable name and self-defined \textit{struct} with the identifier ``\_<funcN>\_'', ``\_<varN>\_'' and ``\_<typeN>\_'' for normalization, where N is the index to denote $\mathrm{N}\-\mathrm{th}$ position. For example, given three variables ``a'', ``b'' and ``c'' in $m_{\mathrm{err}}$, we replace them with ``\_<var1>\_'', ``\_<var2>\_'' and ``\_<var3>\_'' correspondingly.

The processed error message will be fed to the network as a part of the input for the learning module. 
\revise{Normalization can greatly reduce the vocabulary size of the model and has proven to be effective for software vulnerability detection~\cite{li2018vuldeepecker, zou2019mu}.}
\revise{It is worth mentioning that we retain the names of these identifiers in a mapping table and will recover them after the repair is completed. }

\vspace{-2mm}

\subsection{Context Analyzer}\label{sec:context}
As shown in Figure~\ref{fig:intro_example}, the context (line 11) of the error statement (line 9) could reflect the variable ``A'' is a pointer rather than an integer.
However, existing works~\cite{ahmed2021synfix, yasunaga2020graph, mesbah2019deepdelta} usually ignore the context of each statement in learning, which could provide valuable information to program repair.
We propose a context analyzer to extract the context (\ie, $c_i$) of the statement (\ie, $l_i$) in a broken program (\ie, $p$) and take it as part of the input for the enhancement. 

The extraction procedure is presented in Algorithm~\ref{alg:context}. Specifically, we define the input as a program text $p$ and a list of dictionaries $L$, where each dictionary consists of one statement $l_i$, the empty lists of ``vars\_declare'' and ``vars\_use'' for $l_i$ and a dictionary that stores the context for $l_i$. The length of the list $L$ is equal to the number of lines for a program $p$. 
We first design a lexical analyzer (\ie, function ANALYZER) to take $p$ as input and outputs three sets, which are variable names (var\_set), function names (func\_set) and type names (type\_set) respectively. We analyze the token from the union of these sets to obtain its attribute (declaration or usage) and append it into a list of vars\_declare and vars\_use from line 2 to line 8. The function IS\_DECLARE is designed by analyzing the token. If it is a variable/function name or some types come before it, such as ``Integer'' or ``Float'', we believe this token is the declaration and append it into vars\_declare. Otherwise we append it to vars\_use. Similarly, if the token is a type name and followed by the \textit{``struct''} or \textit{``typedef''}, we also append it to vars\_declare. Otherwise, it is appended to vars\_use. Once we have the attribute of a token in the statement, we then extract the context. On one hand, for a token in the list of vars\_use, we retrieve its nearest declaration statement and construct a list of declaration statements about all tokens from vars\_use by the function GET\_DECLARE\_LINES. On the other hand, for the declared token, we also retrieve its nearest usage statement. Since the declared token is usually introduced by the expression such as ``int a = b'', where ``a'' is the declared token and ``b'' is the usage token, we also retrieve the nearest usage statement for ``b'' and combine it with the usage statement of ``a'' to construct a list of usage statements about all tokens from vars\_declare by the function GET\_USE\_LINES. Last, we concatenate the declare context (line[`context'][`declare']) and use context (line[`context'][`use']). We further remove the duplicate and sort them by the order of the original program $p$, then take it as the context $c_i$ for statement $l_i$.

\begin{algorithm}[t]
\small
\caption{Context Analyzer} 
\label{alg:context}
\KwIn{p: program;
      L: List[
            \\
            \hspace{1cm} \{ 
            \\ \hspace{1.5cm}   statement: string; \\ \hspace{1.5cm} vars\_declare: []; \\ \hspace{1.5cm} vars\_use: []; \\ \hspace{1.5cm} context: \{'declare':[], 'use':[]\}
            \\ 
            \hspace{1cm} \}
            \\ ];
      } 
\KwOut{L}
\SetKwFunction{FMain}{get\_context}
\SetKwProg{Fn}{Function}{:}{}
    var\_set, func\_set, type\_set = \textsc{analyzer}(p) \\
    \ForEach{line $\in$ L}{%
      \ForEach{token $\in$  var\_set $\cup$ func\_set $\cup$ type\_set }{
            \If{ token $\in$ line['statement']}{
                \If{\textsc{is\_declare}(line, token)}{
                    line['vars\_declare'].append(token)
                }
                \Else{
                    line['vars\_use'].append(token)
                }
            }
        }
    }
    \ForEach{line $\in$ L}{%
        line['context']['declare'] = \textsc{get\_declare\_lines}(p, line['vars\_use']) \\
        line['context']['use'] = \textsc{get\_use\_lines}(p, line['vars\_declare'] $\cup$  line['vars\_use'])
    }
\end{algorithm}
\section{Program Repair}\label{sec:repair}
\begin{figure}[t]
     \centering
     \includegraphics[width=1\linewidth]{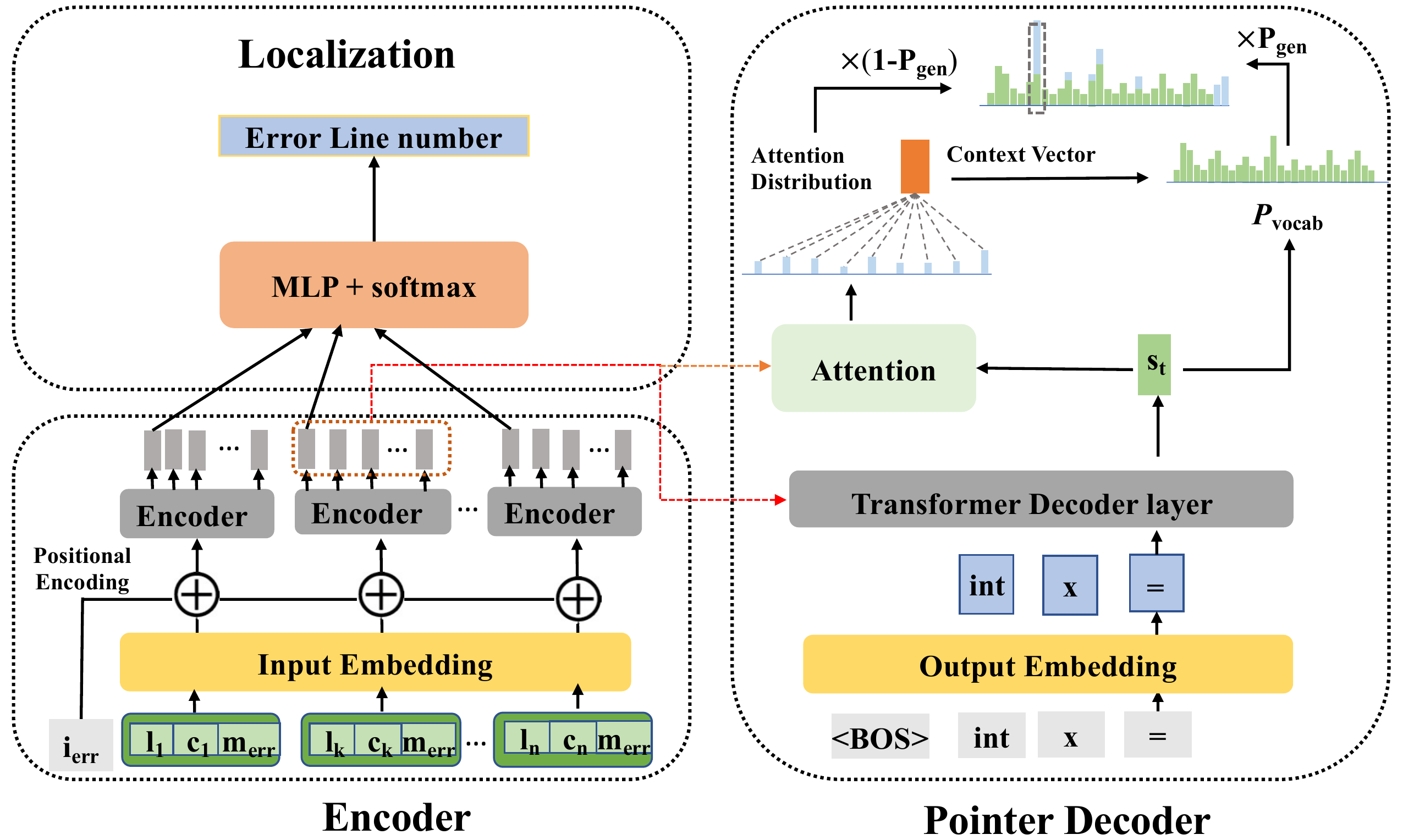}
     \caption{The model architecture of \tool.}
     \label{fig:arch}
     \vspace{-6mm}
\end{figure}

In this section, we introduce the model architecture of \tool, which is shown in Figure~\ref{fig:arch}. It is based on the Transformer architecture and consists of three parts: Transformer-based encoder to encode a broken program to obtain the vector representation of each statement; a fully connected forward neural network (MLP) to locate the broken line, and a pointer decoder to generate a correct statement for fixing.

\subsection{Encoding Broken Programs}\label{sec:encoder}
Through Section~\ref{sec:parsing}, we obtain the compiler feedback $(i_{err}, m_{err})$ of the broken program $p$ and the context $c_i$ for each statement $l_i \in p$. To learn the representations, we directly adopt the Transformer encoder~\cite{vaswani2017attention} for encoding. Specifically, for each statement $l_i$ with its context $c_i$ and the error message $m_{err}$, we construct the input $s_i$ in a format of (<BOS>, $l_i$, <sep>, $c_i$, <sep>, $m_{err}$, <EOS>), and feed it to the Transformer encoder to learn the input representation $\boldsymbol{H}_i \in \mathbb{R}^{m \times d}$, where $m$ is the total number of tokens for the input $s_i$ and $d$ is the dimension length. The calculation can be expressed as follows:
\begin{equation}
\boldsymbol{H}_i = \mathrm{Encoder}(s_i)
\end{equation}

The network architecture of the encoder is almost the same with Vaswani \etal~\cite{vaswani2017attention}, which is composed of a stack of $N$ identical layers and each layer has two sub-layers (the multi-head attention layer and the fully connected feed-forward network). 
The only difference is the positional encoding. We follow DrRepair~\cite{yasunaga2020graph} to add the positional encoding of the line offset with the reported line with error, \ie, $\Delta i = i_{err} - i$, to each token embedding in $s_i$. 

\subsection{MLP for Localization}
By the Transformer encoder in Section~\ref{sec:encoder} for encoding, we obtain each sequence representation $\boldsymbol{H}_i$, where $i \in \{1,2,\cdots,n\}$ and $n$ is the total lines of a broken program. To locate the line of the error statement (\ie, $k$), we turn this localization problem into a classification task. Specifically, we extract the vector of $\boldsymbol{H}_i$ at the symbol ``<BOS>'' (\ie, position ``0'') as the aggregated sequence vector $\boldsymbol{h}_i$ to represent the sequence $s_i$, which is similar to CodeBERT~\cite{feng2020codebert} and use the softmax function with two fully connected layers to determine whether each statement is erroneous or not according to the predicted probability. The loss function $\mathcal{L}_\mathrm{loc}$ can be expressed as follows:

\begin{equation}
    \mathcal{L}_\mathrm{loc} = -\mathrm{log} \frac{\mathrm{exp}(\boldsymbol{h}_k)}{\sum_{i=1}^n \mathrm{exp}(\boldsymbol{h}_i)}
\end{equation}
where $k$ is the location of the error line in the broken program $p$ and $n$ is the total number of lines of $p$.

\subsection{Pointer Decoder for Fixing}
The localization module helps \tool to locate the error statement in a broken program, we further add a decoder to generate a fixed statement for repair. We adopt the transformer decoder and further add pointer mechanism to copy tokens from the input sequence to overcome the out-of-vocabulary (OOV) issue and improve the accuracy of fixing. Specifically, given the output representation $\boldsymbol{H}_k \in \mathbb{R}^{m \times d}$ of the encoder for the broken statement (<BOS>, $l_k$, <sep>, $c_k$, <sep>, $m_{err}$, <EOS>), where $m$ is the sequence length, at each step $t$, we utilize the Transformer decoder~\cite{vaswani2017attention} to receive the word embedding of the previous word and output the hidden states $\boldsymbol{s}_t$. Furthermore, to compute a probability distribution over the input sequence to tell the decoder where to attain to generate the next word, we compute the attention distribution between $\boldsymbol{s}_t \in \mathbb{R}^{d}$ and $\boldsymbol{H}_k \in \mathbb{R}^{m \times d}$, which can be expressed as follows:
\begin{equation}
    \boldsymbol{a}^t = \mathrm{softmax}(\frac{\boldsymbol{H}_k \boldsymbol{s}_t}{\sqrt{d}})
\end{equation}
where $\boldsymbol{a}^t \in \mathbb{R}^{m}$ and $d$ is the dimension length. Then the attention distribution is used to produce a weighted sum of the encoder hidden states (\ie, the context vector):
\begin{equation}
    \boldsymbol{h}^*_t = \sum_i\boldsymbol{a}_i^t\boldsymbol{h}_i
\end{equation}
where $\boldsymbol{h}_i$ denotes $i$-th vector in $\boldsymbol{H}_k$. The context vector is concatenated with the decoder state $\boldsymbol{s}_t$ and produce the vocabulary distribution $P_\mathrm{vocab}$:
\begin{equation}
    \label{eq:vocab}
    P_{\mathrm{vocab}} = \mathrm{softmax} (\boldsymbol{V'}(\boldsymbol{V}[\boldsymbol{s}_t; \boldsymbol{h}^*_t] + \boldsymbol{b}) + \boldsymbol{b'})
\end{equation}

However, Eq~\ref{eq:vocab} could only produce the token from the vocabulary set and the Out-of-vocabulary (OOV) issue, which means that the token is in the input sequence but out of the vocabulary set due to the limited vocabulary length, cannot handle. To address this limitation, similar to See~\cite{see2017get}, we incorporate the pointer mechanism to allow the network to copy words by pointing and generate words from a fixed vocabulary. Specifically, the generation probability $p_{\mathrm{gen}} \in [0,1]$ for each step $t$ is calculated from the context vector $\boldsymbol{h}^*_t$, the decoder state $\boldsymbol{s}_t$ and the decoder input $\boldsymbol{x}_t$:

\begin{equation}
    p_{\mathrm{gen}} = \sigma (\boldsymbol{w}^T_{h^*} \boldsymbol{h}^*_t + \boldsymbol{w}^T_s \boldsymbol{s}_t + \boldsymbol{w}^T_x \boldsymbol{x}_t + b_{\mathrm{ptr}})
\end{equation}
where $\boldsymbol{w}_{h^*}$, $\boldsymbol{w}_s$, $\boldsymbol{w}_x$ and $b_{\mathrm{ptr}}$ are learnable parameters and $\sigma$ is the sigmoid function. $p_{\mathrm{gen}}$ is used to choose between generating a token from vocabulary or copying directly from the input sequence. Over an extended vocabulary set, that combining the original vocabulary set with the tokens from the input sequence, the probability distribution is expressed as follows:
\begin{equation}
    P(w) =p_{\mathrm{gen}}P_{\mathrm{vocab}}(w) + (1 - p_{\mathrm{gen}}) \sum_{i:w_i=w}\boldsymbol{a_i}^t
\end{equation}

The loss function for the fixing (\ie, $\mathcal{L}_\mathrm{gen}$) can be expressed as follows:
\begin{equation}
    \mathcal{L}_\mathrm{gen} = -\frac{1}{T} \sum_{t=0}^T \mathrm{log} P(w_t^*)
\end{equation}
where $w_t^*$ is the target word for timestep $t$ and $T$ is the length of the whole sequence. During the training phase, we directly add the loss values of the location model and the fixing model for training: 
\begin{equation}
\mathcal{L} = \mathcal{L}_\mathrm{loc} + \mathcal{L}_\mathrm{gen}
\end{equation}

\section{Evaluation Setup}\label{sec:eval}
In this section, we first introduce the used datasets for different approaches, then briefly introduce the selected state-of-the-art baselines for comparison and the metrics for evaluation. Finally, we present the details about the model configuration of \tool. We aim at answering the following research questions: 
\begin{enumerate}[leftmargin=*,label=\textbf{RQ\arabic*.}]
    \item What is the performance of \tool compared with current existing state-of-the-art approaches?
    \item Is each component (\ie, diagnostic feedback, context and pointer mechanism) in \tool effective to improve the repair accuracy?
    \item Is each type of the perturbation strategies is beneficial for constructing a more diverse dataset and helping the model improve the performance? 
    \item When \tool fails and when it works? An empirical study for investigating the detailed repaired results compared with the state-of-the-art.
\end{enumerate}

\begin{table*}[t]
\small
\caption{The statistics of the constructed dataset.}
\label{tbl-dataset}
\vspace{-2mm}
\begin{tabular}{cccccccccc}
\toprule
\multirow{2}{*}{\textbf{Correct Programs}} & \multicolumn{5}{c}{\textbf{Training set}}                      &       & \multirow{2}{*}{\textbf{Validation set}} & \multicolumn{2}{c}{\textbf{Test set}} \\ \cline{2-7} \cline{9-10} & struct & stmt & decl & tm & im & Total && TRACER & DeepFix \\ \midrule
 37,415 &  461,663         &   778,210        &    261,944       &  274,366    &    45,074 &   1,821,275    &      2,000    &       3,674       &      6,971         \\ \bottomrule
\end{tabular}
\vspace{-4mm}
\end{table*}

\vspace{-4mm}
\subsection{Datasets}
In the evaluation, we corrupt the correct programs (in total 37,415) from DeepFix~\cite{gupta2017deepfix} and obtain a total number of 1,821,275 synthetic programs for model training. 
\revise{We conduct a strict deduplication process based on code text similarity\cite{allamanis2019adverse} to remove the same samples between the training data and testing data.}
\revise{To construct a validation set, we randomly select 2000 samples from TRACER's training set (17,688 in total) for validation. }
We separately evaluate the performance of the trained model on the testset of DeepFix, which has 6,971 broken programs without ground-truths, and TRACER that contains 3,674 single-line error programs with the provided single-line ground-truths for a comprehensive evaluation. The statistics of the dataset are presented in Table~\ref{tbl-dataset}. Since the broken programs in the testset of DeepFix may contain errors in multiple lines, 
we apply \tool iteratively until the program passes the compilation, or the tries exceed the maximum limit of 5. 

\vspace{-2mm}

\subsection{Baselines}\label{sec:eval:baseline}
\noindent \textbf{DeepFix}~\cite{gupta2017deepfix}. DeepFix firstly proposes to adopt the sequence-to-sequence model for fixing programming errors and it concatenates the line number with the line statement as the input for RNNs with the attention mechanism to generate the error line number and the fixed statement. It further designs an iterative strategy to fix multiple errors in a program and the acceptance standard for one line fixing is whether the updated program can yield less error messages than the input program by the compiler. Furthermore, DeepFix also releases a dataset that has been widely used for the evaluation in the follow-up related works for repairing programming errors. 

\noindent \textbf{RLAssist}~\cite{gupta2019deep}. RLAssist proposes a programming language correction framework based on reinforcement learning, which allows an agent to mimic human actions for text navigation and editing. Specifically, by a trained agent, it allows a set of navigation and edit actions to fix a program. The experimental results proved its superiority against Deepfix.  

\noindent \textbf{SampleFix}~\cite{hajipour2021samplefix}. SampleFix proposes a deep generative model to automatically correct programming errors by learning a distribution over potential fixes. A deep conditional variational autoencoder~\cite{sohn2015learning} is used to sample the fixes for an erroneous program. Furthermore, a novel regularizer is proposed to encourage the model to generate diverse fixes. The experimental results on the DeepFix dataset have confirmed the effectiveness of the proposed architecture. 

\noindent \textbf{MACER}~\cite{chhatbar2020macer}. Since the source code of TRACER~\cite{ahmed2018compilation} is not public and we utilize a follow-up work Macer from the same research team, which has confirmed its superiority over TRACER and been made public. Specifically, MACER conducts a code abstraction procedure and formulates this problem as a classification task by predicting the repaired type in a limited repair classes and applies the predicted repairs at the predicted location. Then, it recovers code abstraction and compiles the fixed program for evaluation. The performance on the DeepFix dataset and TRACER dataset confirms the improvement over TRACER.

\noindent \textbf{DrRepair}~\cite{yasunaga2020graph}. DrRepair incorporates the diagnostic feedback produced by the compiler for a broken program into a designed model and obtains significant improvements against the previous works. Specifically, DrRepair constructs a program-feedback graph to build the relations between a broken program and the feedback. Then model architecture consists of the bidirectional LSTMs~\cite{hochreiter1997long} to learn the statement dependencies and the graph attention network~\cite{velivckovic2017graph} to capture the relations between program and feedback. Furthermore, to construct a large scale dataset for pre-training, DrRepair proposes a program corruption procedure to corrupt correct programs from DeepFix. The extensive experimental results on the DeepFix dataset and SPoC dataset prove that DrRepair could achieve the state-of-the-art performance. In our paper, we compare our approach with DrRepair and its alternative without pretrain (\ie, DrRepair w/o pretrain). 

For DeepFix, RLAssist and SampleFix, 
we directly get the reported values in their original papers. For MACER, we utilize the official released model to test the performance on the TRACER testset and DeepFix testset. For DrRepair and \tool, we separately train the model using the DrRepair-released dataset and our constructed dataset. In addition, in terms of full repair metric on the DeepFix testset, DrRepair sets the beam size to 50 to generate 50 programs for a broken program to test whether this broken program can be fixed. However, by our analysis, we find that the time cost is heavy when setting beam size to 50 and it costs nearly 5 hours for a complete generation process on the DeepFix testset. Considering time and efficiency cost, we set beam size to 5 for DrRepair and \tool for fair comparison.

\begin{table*}[!ht]
    \small
	\caption{The experimental results compared with the baselines where the reported values are in percentages and the values with the marker $*$ denote these values are taken from the corresponding papers directly and the marker - denotes the unreported metrics on the specific testset.}
	\label{tbl-baselines}
	\begin{tabular}{c|c|ccc|c|c}
		\hline
		\multicolumn{1}{c|}{\multirow{3}{*}{\textbf{Model}}} & \multicolumn{5}{c|}{\textbf{TRACER Testset}}                                                                                                                                 & \textbf{DeepFix  Testset}                    \\ \cline{2-7} 
		\multicolumn{1}{c|}{}                       & \multicolumn{1}{c|}{\multirow{2}{*}{\textbf{Single Localize}}} & \multicolumn{3}{c|}{\textbf{Single Repair}}                                                 & \multirow{2}{*}{\textbf{Full Repair}} & \multirow{2}{*}{\textbf{Full Repair}} \\ \cline{3-5} &    & \textbf{Acc@1} & \textbf{Acc@5}  & \textbf{Acc@10} &  &   \\ \hline
		DeepFix      &  -     & -      &    -    &  -         &         -                     & 27.00*                             \\
		RLAssist    &  -      & -      &    -    &  -         &     -                         &  26.60*                            \\
		SampleFix   &  -      & -      &    -    &  -         &     -                         &  45.30*                            \\ \hline
		MACER   &    31.57                &   10.34      &   16.55   &   38.32      &    26.08          &  56.40       \\
		DrRepair\_ori         & 84.98 &  46.24 & 57.73   & 60.13 &  72.66   & 62.13   \\ 
		DrRepair\         & \textbf{86.72} &  48.56 & 60.23   & 62.28 &  77.11   & 63.87    \\ \hline
		\emph{TransRepair}\_ori      & 80.19  & 44.47 &  58.57     &  63.39   & 78.77 & 66.71       \\ 
		\tool      & 83.21  & \textbf{49.65} &  \textbf{61.27}     &  \textbf{65.08}   & \textbf{82.81} &\textbf{68.53}       \\ \hline
	\end{tabular}
\vspace{-5mm}
\end{table*}

\vspace{-3mm}

\subsection{Metrics}
\revise{We evaluate our approach against other baselines in the metrics of single localize, single repair and full repair accuracy. Since TRACER's testset provides the ground-truths of single-line erroneous program (\ie, each broken program has its correct counterpart), we could use all these metrics for evaluation.
However, we only utilize full repair accuracy for the DeepFix testset since it is without ground-truths.}

\noindent \textbf{Single Localize.} It defines the accuracy of localizing a single error statement in a single-line error program in the TRACER testset.

\noindent \textbf{Single Repair.} It is used to evaluate if the generated statement is exactly matched with the ground-truth associated with a broken statement. In this setting, we assume that the error statement is known and we do not need a localization module for localizing an error statement. We use Acc@k to calculate the percentage of the correct results existed in the top-k returned results. Specifically, we adjust the beam search size equal to $k$ to return $k$ results for a broken program and we set $k$ to 1, 5, 10 to evaluate the accuracy of the generated statement in TRACER where each sample has a ground-truth for calculation. 

\noindent \textbf{Full Repair.} It is designed to evaluate the ability of different approaches on fixing a broken program, which consists of localizing an error statement and further fixing it. Furthermore, it is calculated in the percentage of the generated program that could pass the compiler in success. We utilize full repair accuracy in both the TRACER testset and DeepFix testset for evaluation.  
\revise{It is noted that the metric ``Full Repair'' may have limits for evaluation considering the scenarios when the erroneous lines are simply removed rather than correctly edited. It is used here because: 1) the Deepfix dataset has no ground-truths, so we resort to full repair to evaluate the repair performance. Meanwhile, we avoided deleting the entire line which may incur dramatic changes to code semantics.  2) These metrics have also been widely used in~\cite{ahmed2018compilation,chhatbar2020macer,yasunaga2020graph}, with which we can compare with prior studies directly. But we will explore more better metrics in future.}

\vspace{-2mm}

\subsection{Model Configuration}
\tool consists of 5 identical layers for the Transformer encoder and decoder, each layer has 8 heads to learn different subspace features. We select the tokens with the frequency greater than 1 in the training set for constructing our vocabulary set. The word dimension is set to 256 with the positional encoding equals to 50 for the embedding. The optimizer is selected with Adam~\cite{kingma2014adam} with an initial learning rate of 0.0001 and batch size of 25. We set the dropout to 0.1 and gradient clipping to 10. All hyper-parameters are tuned on the validation set. The model is trained on a Intel(R) Xeon(R) server with 8 cores, which equips Nvidia 3090 with 24G memory and 2 Nvidia TITAN X with 12G memory and the training process costs around 30 hours. 
\vspace{-2mm}
\section{Evaluation Results}\label{sec:results}
In this sections, we present the experimental results in light of research questions.
\vspace{-2mm}
\subsection{RQ1: Comparisons with Baselines}\label{sec:eval:comparison}
We compare \tool with some existing approaches, specifically the row of ``$\{*\}$\_{ori}'' indicates the model $\{*\}$ trained on the original training set that DrRepair released. The experimental results are presented in Table~\ref{tbl-baselines}.

Among different baselines, we find that DrRepair could achieve the best performance on both TRACER and DeepFix testset, which is in line with the perception that DrRepair is current state-of-the-art approach for repairing program syntax errors. Furthermore, we can observe that \tool could obtain higher single repair and full repair accuracy than DrRepair when fixing a training set to train (\ie, the original training set that DrRepair uses or our corrupted training set), which illustrates the superiority of our approach in program repair against DrRepair. However, we also find that the accuracy of single localize of \tool is lower than DrRepair on the TRACER testset, we conjecture that it is caused by the localization requires the exact match to the error line, which is harder for \tool (Transformer-based) to achieve higher performance compared with DrRepair (LSTM-based).
However, the requirement for generating a statement to replace the error statement to pass the compilation is relatively easier for \tool since the Transformer is more powerful than LSTMs in generating a target sequence even in adverse condition when Transformer has poor ability to accurately localize an error statement. The more powerful generation ability of transformer can be further enhanced by comparing the results of \tool and DrRepair on the single repair accuracy and this metric is used to evaluate the generated statement is exactly matched with the ground-truth when taking an error statement as the input for the decoder. We can observe that \tool achieves higher single repair accuracy than DrRepair. Hence, the poor localize accuracy may not significantly impact the repair accuracy in our model and we believe that the metric of repair accuracy plays a critical role for program repair. But we also want to investigate the way to improve our localization accuracy and we leave it as our future work. 

In addition, fixing a model (\eg, DrRepair or \tool), we use our constructed training set or the original training set that DrRepair used for training separately.
We could achieve higher repair accuracy on our training set than the original training set that used by DrRepair.
It proves that by our designed perturbation strategies, we can construct a training set that is more in line with the real scenario and this dataset could help the model achieve a better performance.

\noindent\textbf{In summary:} 
\revise{\tool provides higher repair accuracy compared with the state-of-the-art approach DrRepair.} 
We attribute the improvements to the powerful generation ability of the Transformer. Furthermore, by comparing the performance among the training sets that DrRepair used and we constructed, we further confirm that our training set is better for the model to achieve higher performance.

\subsection{RQ2: Ablation study of each component in the network architecture}
We ablate the performance of \tool when removing the specific component in the model architecture and maintaining the others for evaluation. The experimental results on TRACER in terms of single repair accuracy are presented in Table~\ref{tbl-ablation-component}, where w/o denotes the removed component in \tool and the model configuration is the same as \tool for fair comparison. 

As shown in Table~\ref{tbl-ablation-component}, we can find that the diagnostic feedback plays a critical role in improving the performance and removing it degrades the accuracy significantly. This shows that the diagnostic feedback could supplement some valuable information such as the error line and error message, although in many cases, this information may be inaccurate, it could still contribute the model to achieve higher accuracy when incorporating this part of information. Furthermore, we can observe that the context is also important in improving the performance. Since the context of a statement could reduce the difficulty for the model to learn this statement semantics (See an example in Figure~\ref{fig:intro_example}, ignoring the context will limit the repair accuracy. The pointer mechanism could effectively alleviate the out-of-vocabulary issue and without it. The repair accuracy drops from 65.08 to 64.78, which demonstrates that there are some target tokens might be out of vocabulary set. Hence, we incorporate the pointer mechanism into the Transformer decoder can mitigate this issue and further improve the performance. Overall, from Table~\ref{tbl-ablation-component}, we can conclude that when combing all of these components, \tool could achieve the best repair accuracy.

\noindent\textbf{In summary:} 
The diagnostic feedback plays a critical role in improving the repair accuracy, however the contextual information and the pointer mechanism is also beneficial for the improvement and when incorporating all of components, \tool could achieve the best performance.

\begin{table}[t]
\small
\caption{The ablation results of \tool, where w/o denotes the removed component. }
\label{tbl-ablation-component}
\begin{tabular}{cccc}
\toprule
\multirow{2}{*}{\textbf{Model}} & \multicolumn{3}{c}{\textbf{TRACER Testset}} \\ \cline{2-4} 
                       & \textbf{Acc@1}     & \textbf{Acc@5}     & \textbf{Acc@10}     \\ \hline
w/o feedback       &   45.67        &   58.74        &  62.38          \\
w/o context        &   47.09        &   59.93        &  64.07          \\
w/o pointer        &   47.00        &   60.86        &  64.78          \\ \hline
\tool                  &  \textbf{49.65}         &  \textbf{61.27}         &  \textbf{65.08}          \\ \bottomrule
\end{tabular}
\vspace{-5mm}
\end{table}

\begin{table}[t]
\small
\caption{The ablation results by removing one type of perturbation strategies to construct the training set for learning.}
\label{tbl-ablation-strategy}
\begin{tabular}{cccc}
\toprule
\multirow{2}{*}{\textbf{Model}} & \multicolumn{3}{c}{\textbf{TRACER Testset}} \\ \cline{2-4} 
                       & \textbf{Acc@1}     & \textbf{Acc@5}     & \textbf{Acc@10}     \\ \midrule
w/o struct       &   48.06       &   60.51    &  64.43        \\
w/o stmt        &    47.23       &   59.56      &   62.66       \\
w/o decl        &    46.57      &   58.49      &   62.40     \\ 
w/o tm        &   47.88      &    59.51       &   63.37       \\
w/o im        &   47.35      &    59.73      &   63.06     \\\hline
\tool                  &  \textbf{49.65}         &  \textbf{61.27}         &  \textbf{65.08}          \\ \bottomrule
\end{tabular}
\vspace{-6mm}
\end{table}

\subsection{RQ3: Ablation study of perturbation strategies in dataset construction}
In Section~\ref{sec:synthesis}, we design a set of 5 perturbation strategies and sample 1-5 strategies to corrupt a correct program and construct a training set for \tool to learn. In this RQ, we also investigate the effect of each perturbation strategy in building the training set. Specifically, we remove one type of perturbation strategies and maintain the others to build a new training set where the total number of samples in this training set is equal to the original one. Then we train our model on the newly constructed training set with the same model configuration as the original to compare the performance, and the experimental results are presented in Table~\ref{tbl-ablation-strategy}. 

We can see that each perturbation strategy is effective in constructing a more diverse training set. When combing all to build a training set, we could achieve the best performance. Specifically, when the training set is constructed without the structure strategy (\ie, the training set has no samples with the type of structure error) has the lowest drop in repair accuracy compared with other strategies.
It depicts that the structure error type has the least contributions in constructing a diverse training set to help the model obtain higher repair accuracy. We infer that it is caused by the difficulty in fixing this type of errors. The defined operations only modify punctuators such as ``\{'', ``\}'' in a correct program and these punctuators have no semantic information for a program compared with other types of corrupted operations in Table~\ref{tbl-statistics}, which involves modifying the variable names to synthesize other error types. 
Hence, the model is difficult to learn effective patters for structure errors and removing this type of data in the training set cannot lead the model have a significant impact on the repair accuracy. Furthermore, we can see that removing the data that have the ``variable declaration (decl)'' errors in the training set, the repair accuracy decreases significantly and it demonstrates that adding samples with this type of error could be beneficial for the model learning. We believe that the improvement is due to
the designed context analyzer (see Section~\ref{sec:context}), which could extract the contextual information for these variables and it is significantly beneficial for the model to learn effective repair patterns.

\noindent\textbf{In summary:} 
Each type of perturbation strategies is beneficial in constructing a diverse training set. When combining them together and apply them to corrupt correct programs for building the training set, we could obtain the best repair accuracy.

\subsection{RQ4: When \tool fails and when it works?}
\begin{figure}[!t]
     \centering
      \includegraphics[width=1\linewidth]{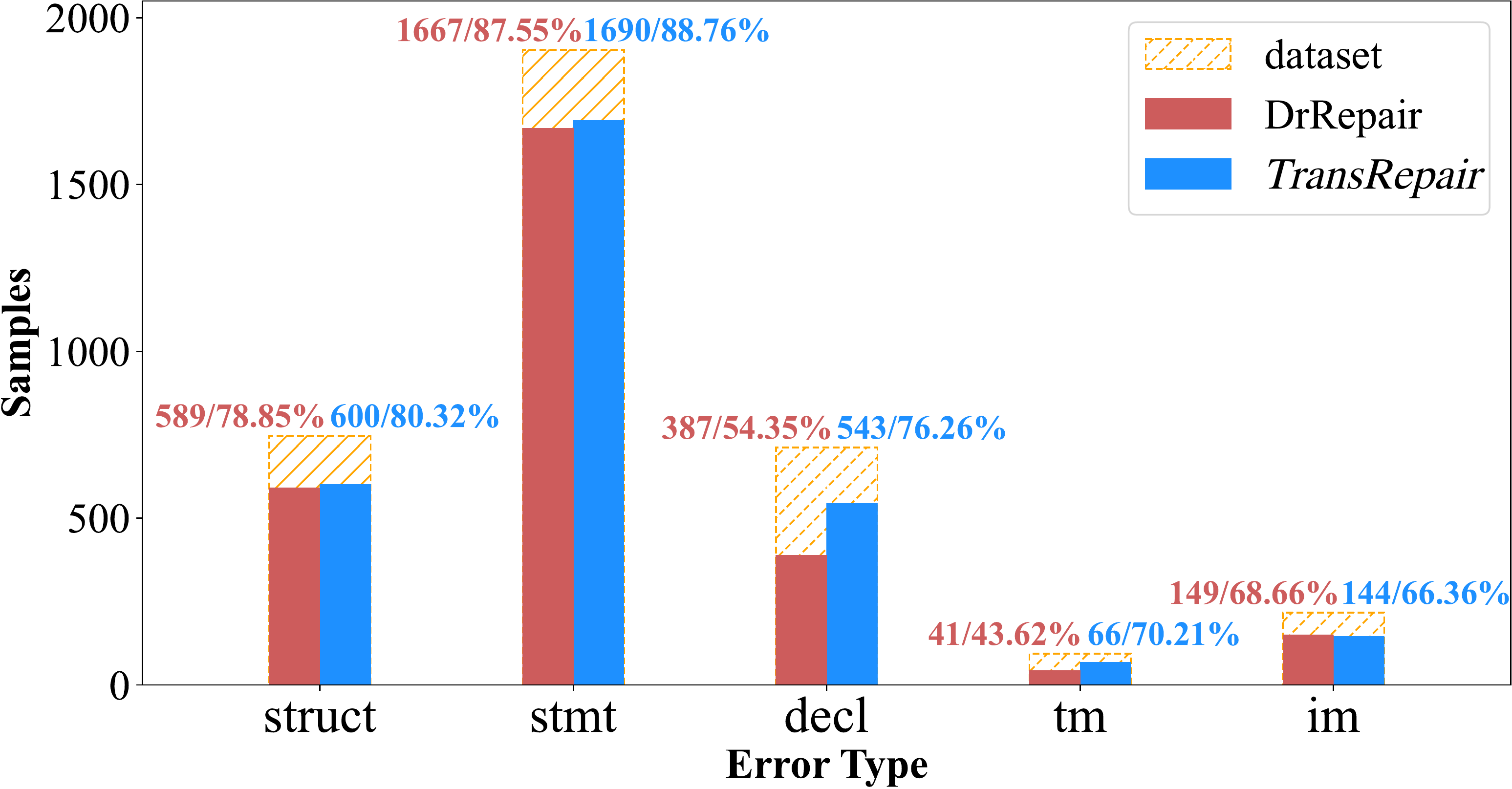}
     \caption{\revise{The comparison results for the number of repairs between DrRepair and \tool.}}
     \label{fig:number}
     \vspace{-5mm}
\end{figure}

We conduct a statistical analysis further to compare the repaired results between \tool and DrRepair.
Both models are trained on our constructed dataset and tested on the TRACER testset to verify model's ability to fix different types of program errors. 
The statistical results are presented in Figure~\ref{fig:number}, where the number besides the rectangle is the total number of fixes and the ratio of the number of fixes to the total number of this type errors. More details on the repair efficacy for each concrete error pattern can be found on our website~\cite{website}. 

\begin{figure}[t]
\centering
\centering
     \includegraphics[width=1\linewidth]{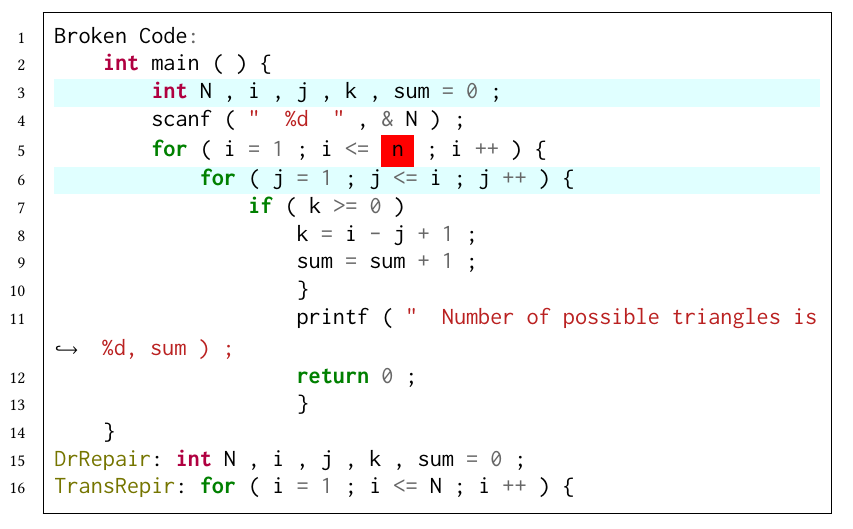}
\caption{The broken program with the generated statement by DrRepair and \tool for fixing.}
\label{fig:results_example}
\vspace{-6mm}
\end{figure}

As illustrated in Figure~\ref{fig:number}, we find that \tool is excellent in fixing the errors of ``variable declaration (decl)'' and ``type mismatch (tm)''  and has a slight improvement in fixing the errors of ``structure (struct)''  and ``statement (stmt)'' while is slightly inferior to fix the error of ``identifier misuse (im)'' compared with DrRepair. 
We conjecture that DrRepair could fix more ``identifier misuse (im)'' errors due to the constructed program feedback graph to capture the variable relations.
However, we can also get a competitive performance by the powerful Transformer without the need of the constructed graph. For the other four errors that \tool could fix better than DrRepair, we attribute the improvement to the used context in helping the model capture the error statement patterns. Especially for the error type of variable declaration, the context information around the error statement is critical to reveal the root cause. Here we present one example with the generated results by \tool and DrRepair in Figure~\ref{fig:results_example} for better illustration. 
It shows that the error is due to the variable ``n'' is not defined at line 5 and its contexts are highlighted in blue at line 3 and line 6. We encode the context (\ie, line 3 and line 6) for this error statement could help \tool generate a correct statement ``for ( i = 1 ; i <= N ; i ++ ) \{'' for the fixing, while due the lack of the context, DrRepair fails to generate a correct statement to repair this error.  

\noindent\textbf{In summary:} 
Generally, \tool is competitive in fixing the type mismatch error compared with DrRepair, however on other four errors, it could achieve better performance, we attribute the improvement to the utilized context for learning.  

\vspace{-2mm}
\section{Threats to validity}\label{sec:discuz}
\textbf{Internal validity.} One of the threats to validity is the hyper-parameter setting for our approach. We tune our model on the validation set and select the best model based on the repair accuracy and use it for testing. We will explore more hyper-parameters for our approach. Another threat lies in our implementations of the broken code synthesis, context analyzer and model implementation. To reduce this threat, the authors carefully check the correctness of the implementation. We will make our code and the constructed dataset public for further investigation. 

\noindent \textbf{External validity.} The external threats to validity include the selected datasets, the evaluation of the baselines and the evaluation metrics. In terms of dataset, the training set is constructed from the DeepFix dataset, which is a popular C programming language dataset for program repairing. 
\revise{We only use three operations (\ie, ADD, DEL and REP) as the building blocks for code mutation. It is intriguing to explore more complex transformation strategies such as multiple operations with logical relations.}

Furthermore, we select two testsets (\ie, TRACER and DeepFix) and they are both on C language for evaluation. 

\revise{We admit that there are some works~\cite{mesbah2019deepdelta} for other languages like Java but TransRepair cannot be directly used for these languages. We will extend our approach with adaptive language analyzers for other languages in future.}
\revise{Additionally, programs in a more complex system may encounter varying compilation errors considering dependent libraries, templates and generics. Therefore, it may degrade the repair performance of TransRepair, which can be to some extent mitigated by involving more complex programs during the training. Our approach remains effective for AI programming such as automated code completion, we can use our approach to repair the generated programs by AI models.}
In terms of baselines, for DeepFix, RLAssist and SampleFix, we report the values from the original paper and we believe these reported values are the best for their approaches. For MACER, we also believe the released model is the optimal and for DrRepair, we only adjust the beam search size to 5 and keep the other settings same with the default configuration for reproduction. The default hyper-parameters of DrRepair on our constructed dataset may not be optimal, however on its original dataset that DrRepair uses (the configuration should be optimal), our experiments prove that our approach outperforms it significantly (see Table~\ref{tbl-baselines}). As for evaluation metrics, we follow DrRepair~\cite{yasunaga2020graph} and utilize the single localize accuracy, single repair accuracy (Acc@1 in our work), full repair accuracy for evaluation. We further add Acc@5 and Acc@10 for a comprehensive evaluation.

\vspace{-2mm}
\section{Related Work}\label{sec:related}
There is a line of works on automated program repair for compilation errors and context-aware program repair. We also briefly introduce some works that use deep learning techniques for different software engineering applications. 
\vspace{-2mm}
\subsection{Automated Compilation Error Repair}

Over the past years, automated program repairs for compilation errors have attracted widespread attention. DeepFix~\cite{gupta2017deepfix} applied a RNN-based encoder-decoder framework to repair program syntax errors on C programming language. RLAssist~\cite{gupta2019deep} is a follow-up work after the DeepFix. It attempted to use deep reinforcement learning to achieve better repair accuracy. 
TRACER~\cite{ahmed2018compilation} also adopted RNN-based model to repair the syntax errors and the follow-up work MACER~\cite{chhatbar2020macer} formulated this problem as a classification task. Hajipour \etal proposed SampleFix~\cite{hajipour2021samplefix}, which applied a deep generative model to fix programming errors automatically. These works just utilize the program for the repair, while some external information such as the diagnostic feedback is ignored. To supplement this part of information, SynFix~\cite{ahmed2021synfix} proposed to incorporate the compiler diagnostics from JavaC with the pre-trained RoBERTa for improvement. Yasunaga \etal proposed DrRepair~\cite{yasunaga2020graph}, which constructed a graph between the diagnostic feedback and the broken program and took them as the input of a self-supervised learning framework to repair the errors. 
Compared with these works, we craft high-quality training data that is in line with the real scenario and made this well-designed dataset public for further studies. Furthermore, we propose a Transformer-based program repair model with pointer mechanism, which incorporates the broken program and the context and diagnostic feedback to improve repair accuracy.   

\vspace{-3mm}
\subsection{Context-Aware Program Program Repair}
Because of the complexity of a broken program, it is hard to accurately capture the program semantics. More researchers attempt to utilize the context as the auxiliary information to enhance the fault localization and program repair for logic errors. Specifically, 
Chilimbi \etal~\cite{chilimbi_holmes_2009} proposed a static analysis approach, namely HOLMES, which determines the root causes of targeted bugs based on the run-time profiling information representing program context.
Wen \etal~\cite{wen_context-aware_2018} proposed a context-aware patch generation approach called CapGen, which leverage several novel prioritization methods to enhance the success rate of automatically generated patch for repair.
Li \etal~\cite{li_dlfix_2020} proposed a context-based code transformation learning approach, namely DLFix, which applied deep learning on automated program repair (APR) without requiring ant hard-coding of bug-fixing patterns.
Lutellier \etal~\cite{lutellier_coconut_2020} proposed a combined neural machine translation (NMT) models based context-aware approach, called CoCoNut, which could work on automatic bug repair in multiple programming languages.
Kim \etal~\cite{kim_effectiveness_2020} proposed ConFix, which is an context-based automatic patch generation approach for buggy programs.
Chen \etal~\cite{seqr}proposed a sequence-to-sequence based tool, namely SequenceR, to repair buggy programs by learning from the buggy context of single line repair from human commits. The main difference between it and ours is that we are focusing on compilation errors other than logic bugs. Inspired by above works, in \tool, we also incorporate context of the error statement for fixing compilation errors.
\vspace{-3mm}
\subsection{Deep Neural Networks for SE Applications}
With the rapid development of AI techniques, more researchers attempt to utilize deep learning techniques for software engineering applications. Compared with traditional software analysis techniques, deep learning techniques aim at learning features automatically from a large amount of data. By training a deep neural network and deploying it to the test phase, the superior performance of these models has been confirmed on different applications. For example, Allamanis et al.~\cite{allamanis2017learning} proposed to construct the program graph and utilized it with Gated Graph Neural Network to learn program semantics for variable misuse detection. Followed by this work, many other works proposed to extract program structures for other applications such as source code vulnerability detection~\cite{zhou2019devign, cheng2021deepwukong}, code summarization~\cite{liu2020retrieval, fernandes2018structured}, deep code search~\cite{liu2021graphsearchnet, ling2021deep},neural program decompilation~\cite{liang2021neutron}. An empirical study~\cite{siow2022learning} is also conducted to illustrate different program structures to the effect of software engineering applications. Recently, more pre-trained models are proposed to learn general code fragment representation for ``code intelligence'' such as CodeBERT~\cite{feng2020codebert} and GraphCodeBERT~\cite{guo2020graphcodebert}. A BART-based pre-trained model CommitBART~\cite{liu2022commitbart} is also proposed for different commit-related applications such as commit message generation~\cite{liu2020atom, jiang2017automatically}, security patch identification~\cite{zhou2021spi, wu2022enhancing}.

\section{Conclusion}\label{sec:concl}
We develop a Transformer-based approach \tool to automatically fix compilation errors in C programs. To craft high quality training data, we spend around 2-man months investigating the compilation errors from 28,965 erroneous programs from two public datasets and the Internet, and then distill 74 error patterns that fall into 5 classes. A data synthesis approach is devised by corrupting correct programs into these errors and finally we obtain 1,821,275 erroneous programs of high diversity. 
\tool is built on top of Transformer that takes as input the broken program, together with its context and the diagnostic feedback. 
It integrates the pointer mechanism to address the out-of-vocabulary code tokens and outputs the localization of the error statement and further provides a fixed version. 
The extensive experiments on two open-source testsets have proved that \tool outperforms the current state-of-the-art both in repair accuracy. 
\vspace{-2mm}
\section{acknowledgment}
We would thank the anonymous reviewers for their valuable comments. IIE authors are supported in part by NSFC (61902395, U1836211), Beijing Natural Science Foundation (No.M22004), the Anhui Department of Science and Technology under Grant 202103a05020009, Youth Innovation Promotion Association CAS, Beijing Academy of Artificial Intelligence (BAAI). This research is also partially supported by the National Research Foundation, Singapore under its the AI Singapore Programme (AISG2-RP-2020-019), the National Research Foundation, Prime Ministers Office, Singapore under its National Cybersecurity R\&D Program (Award No. NRF2018NCR-NCR005-0001), NRF Investigatorship NRF-NRFI06-2020-0001, the National Research Foundation through its National Satellite of Excellence in Trustworthy Software Systems (NSOE-TSS) project under the National Cybersecurity R\&D (NCR) Grant award no. NRF2018NCR-NSOE003-0001, the Ministry of Education, Singapore under its Academic Research Tier 3 (MOET32020-0004). Any opinions, findings and conclusions or recommendations expressed in this material are those of the author(s) and do not reflect the views of the Ministry of Education, Singapore.

\bibliographystyle{ACM-Reference-Format}
\bibliography{main}



\end{document}